\documentclass[preprint,aps,nofootinbib]{revtex4}

\usepackage{epsfig}

\def\bea{\begin{eqnarray}}
\def\eea{\end{eqnarray}}
\def\bec{\begin{center}}
\def\ec{\end{center}}

\def\beq{\begin{equation}}
\def\eeq{\end{equation}}

\begin{document}
\draft
\tighten
\preprint{KUNS-2024}
\preprint{KANAZAWA-06-05}
\title{\large \bf Dynamical realization of democratic Yukawa matrices \\
and alignment of A-terms}
\author{
Tatsuo Kobayashi \footnote{kobayash@gauge.scphys.kyoto-u.ac.jp}$^1$,
Yuji Omura\footnote{ohmura@phys.h.kyoto-u.ac.jp}$^2$ and
Haruhiko Terao\footnote{terao@hep.s.kanazawa-u.ac.jp}$^3$
}
\address{
$^1$Department of Physics, Kyoto University,
Kyoto 606-8502, Japan \\
$^2$Department of Physics, Kyoto University, 
Kyoto 606-8501, Japan \\
$^3$Institute for Theoretical Physics, Kanazawa
University, Kanazawa 920-1192, Japan }
\begin{abstract}
We study realization of the democratic form of Yukawa 
matrices by infrared fixed points.
We investigate renormalization-group flows of 
Yukawa couplings in models with a single Yukawa matrix for 
three families, and up and down-sector Yukawa matrices.
It is found that each model has its certain pattern of 
renormalization-group flows of Yukawa matrices.
We apply them to the charged lepton sector and 
quark sector, and show that realistic results for the 
second and third families are obtained with natural 
initial conditions, e.g. $(V_{MNS})_{23} \sim 2/\sqrt{6}$,
$m_s/m_b \sim V_{cb} \sim m_{\mu}/m_\tau$ and 
$m_c/m_t \sim (m_s/m_b)^{3/2}$.
We also study corresponding A-terms.
The A-terms approach toward the universal form with 
no physical CP-violating phase.
Thus, constraints due to various neutral flavor changing 
processes except for $\mu \rightarrow e \gamma$ are 
found to be satisfied by this dynamics.
In order to suppress the electric dipole moments as well as 
$\mu \rightarrow e \gamma$ sufficiently, more 
alignment of the A-terms with some reason is required.

\end{abstract}
\maketitle

\section{introduction}

Understanding the origin of the hierarchical structure of 
quark/lepton masses and their mixing angles is one 
of important issues in particle physics.
Actually, several types of scenarios have been 
proposed so far.
Among them, the democratic Ansatz on Yukawa 
matrices is an interesting approach 
\cite{Fritzsch:1979zq,Fritzsch:1995dj,Tanimoto:2000fz,Tanimoto:1999pj}, 
in particular for the lepton sector, 
because it can lead to almost bi-maximal mixing angles.

The exactly democratic form of the Yukawa matrix can 
be obtained by a certain flavor symmetry, e.g. $S_3$ symmetry.
However, the exact one by itself is not realistic, 
because it is a rank-one matrix.
Thus,  a certain pattern of small symmetry breaking terms are 
usually added by hand for the purpose 
to realize realistic Yukawa matrices.
However, it seems that any comprehensive explanation 
leading to such pattern of symmetry breaking terms has 
not been given.

Recently, two of the authors have considered dynamical 
realization of the almost democratic Yukawa matrices 
\cite{Kobayashi:2004ha}\footnote{See also \cite{Abel:1998wh}.}, 
that is, each entry of the Yukawa matrix has the same infrared (IR) 
fixed point and the almost democratic Yukawa 
matrix can be realized by dynamics at the IR region 
whatever their initial values are.
Actually, these fixed points correspond to the 
so-called Pendelton-Ross fixed point \cite{Pendleton:1980as}.
The exact IR fixed point is not realistic, because we 
just obtain the exact rank-one Yukawa matrix.
Some region close to the IR fixed point would be interesting, 
and deviations from the fixed point correspond to small breaking
parameters  in the flavor symmetry approach to the democratic Ansatz.
Thus, one of our purposes in this paper is to study more 
about renormalization group (RG) flows of 
Yukawa couplings in models with the above IR fixed points 
leading to the democratic Yukawa matrices.
We shall show that we obtain certain patterns of RG-flows of 
Yukawa matrices, and study 
how much realistic Yukawa matrices 
for the quark and lepton sectors are 
derived from natural initial conditions, 
and which type of fine-tuning is required.

Supersymmetric extension is 
attractive as new physics beyond the standard model.
Within the framework of supersymmetric standard models, 
a flavor mechanism, which derives realistic quark/lepton masses 
and their mixing angles, would affect somehow their 
superpartners, that is, soft supersymmetry (SUSY) breaking 
squark/slepton masses and trilinear couplings, the so-called 
A-terms.
Although superpartners have not been detected yet, 
patterns of  squark/slepton mass matrices and 
A-term matrices are strongly constrained 
by experiments on flavor changing neutral current (FCNC) 
processes \cite{Gabbiani:1996hi}.\footnote{
See also Ref.~\cite{Chankowski:2005jh} and 
references therein.}
That is, FCNCs require almost degenerate  
A-terms and soft scalar masses.
In particular, the constraints due to the 
the $K^0- \bar K^0$ mixing and 
$\mu \rightarrow e \gamma$ decay are severe.
In addition, CP-violating phases of A-terms are also 
strongly constrained by experiments on electric dipole moments (EDMs).
These are the SUSY flavor and CP problems.
A solution for theses problems is to realize the universal 
soft scalar masses and the universal A-terms, 
whose phases are aligned with the phase of gaugino masses, 
e.g. the gluino mass.
In this case, the A-term has no physical CP-violating phase.
Thus, it is quite important to examine these constraints 
on any SUSY flavor mechanism.

So far various flavor symmetries have been also applied to 
suppress the flavor non-universality in SUSY breaking parameters, 
while the mass matrices are also explained by the same symmetries.
As for the democratic mass matrices, the $S_3$ symmetry 
also constrains structure of the SUSY breaking parameters
\cite{Hamaguchi}. 
However, the FCNC or the CP problem is not solved by the 
symmetry alone. Besides, the breaking effects of the flavor 
symmetry on the SUSY breaking parameters are unclear.

It is known that when a model with one flavor has a 
Pendelton-Ross fixed point, the corresponding A-term $A$ as 
well as the sum of soft scalar masses squared also 
has an IR fixed point like $A= -M$, where $M$ is 
the gaugino mass \cite{Lanzagorta:1995ai,Kobayashi:2000wk}.
It is expected that our model with three flavors has 
a similar fixed point, and A-terms are aligned dynamically 
at the IR region \cite{Kobayashi:2000fi}.
Indeed, we shall show IR fixed points of A-term matrices 
are universal $A_{ij}= -M$.
That is a favorable aspect, but the exact IR fixed point 
is not realistic, because we just obtain the 
rank-one Yukawa matrix on the exact IR fixed point as 
said above.
Some region close to the fixed point would be interesting.
Hence, in this paper we study RG-flows of 
Yukawa couplings and A-terms.
We evaluate how much A-term matrices differ from the 
universal form, when Yukawa matrices are somehow realistic.

This paper is organized as follows.
In section II, we review briefly the 
Pendelton-Ross fixed point of Yukawa coupling and 
the corresponding A-term in a SUSY one-flavor model.
In section III, we review briefly 
our three-flavor model, where all entries of a
Yukawa matrix have fixed points.
We study fixed points of the corresponding A-terms, 
$A_{ij}$.
In section IV, we study RG-flow of Yukawa couplings 
such as to obtain a certain pattern of the Yukawa matrix 
by natural initial conditions.
We apply our results to the charge lepton sector.
We also analyze RG-flows of A-terms and examine 
their non-universality.
In section V, we study RG-flows of up and down sector 
Yukawa matrices.
We use $SU(5)'\times SU(5)''$ model as a concrete model.
We show that a certain pattern of up and down sector Yukawa matrices 
are obtained.
We also discuss RG-flows of A-terms and consider 
FCNC and CP-violation constraints.
In section VI, we comment on degeneracy of sfermion masses.
Section VII is devoted to conclusion and discussion.
In Appendix, we give examples with additional couplings, 
where values of fixed points are shifted significantly.

\section{IR fixed point in one-flavor model} 

\subsection{IR fixed point of Yukawa coupling}

First, we briefly review on the Pendelton-Ross IR fixed point 
in the model with a simple gauge group and a single Yukawa coupling, 
which corresponds to the superpotential,
\begin{equation}
W = y \Phi_1 \Phi_2 \Phi_3,
\end{equation}
where $\Phi_i$'s are chiral superfields.
The one-loop RG equations for the gauge coupling $g$ and the Yukawa 
coupling $y$ are obtained  as
\begin{eqnarray}
\label{1-flavor-rg-g}
\mu \frac{d \alpha_g}{d\mu} &=& - b \alpha_g^2, \\
\mu \frac{d \alpha_y}{d\mu} &=& (a \alpha_y -c \alpha_g)\alpha_y, 
\label{1-flavor-rg-y}
\end{eqnarray}
where $\alpha_g \equiv g^2/(8\pi^2)$, 
$\alpha_y \equiv y^2/(8 \pi^2)$, $b$ is the one-loop beta 
function coefficient of the gauge coupling, and 
$a$ and $c$ are group-theoretical constants and both are 
always positive.
Now, let us consider the RG equation of the ratio 
$x=\alpha_y/\alpha_g$,
\begin{equation}
\mu \frac{d x}{d \mu} = [ax - (c-b)]\alpha_gx.
\end{equation}
This equation has a nontrivial fixed point at 
\begin{equation}
x^* = \frac{c-b}{a}.
\end{equation}
The condition $c-b >0$ should be satisfied 
such that this fixed point is physical, i.e. 
$x >0$.

Actually, we can solve this equation exactly, and 
show this fixed point is IR attractive.
However, analysis on linear perturbation around 
the fixed point $x = x^* + \Delta x$ would be useful 
for discussions in the following sections.
Here we study the RG equation for $\Delta x$, 
which satisfies
\begin{equation}
\mu \frac{d \Delta x}{d \mu} = (c -b) \alpha_g \Delta x.
\end{equation}
Thus, this fixed point is IR attractive when $c-b >0$.
Actually, its solution is obtained as 
\begin{equation}
\frac{\Delta x(\mu)}{\Delta x(\Lambda)} = \left( 
\frac{\alpha_g (\mu)}{\alpha_g (\Lambda)} \right)^{{(b-c)}/{b}} 
= \left( 
\frac{\alpha_g (\mu)}{\alpha_g (\Lambda)} \right)^{- ax^*/{b}} .
\end{equation}
Large values of $\alpha_g$ and $(c-b)$ lead to 
stronger convergence toward the IR fixed point.
Asymptotically non-free theories, i.e. $b<0$, are favorable to 
realize strong convergence.

\subsection{Infrared fixed point of A-term}

Next, we review on fixed points of A-terms.
In general, softly broken SUSY models have trilinear 
couplings of scalar components, that is, the so-called 
A-terms.
In the above model, we would have the 
soft SUSY breaking term,
\begin{equation}
h \phi_1 \phi_2 \phi_3,
\end{equation}
where $\phi_i$ are scalar components of 
chiral superfields $\Phi_i$.
Here and hereafter, we use the notation of 
A-term $A=h/y$.

Now let us consider the RG equations of 
the gaugino mass $M$ and the A-term.
It is straightforward to obtain those RG equations 
in a simple model.
However, it is convenient to use the spurion formalism
to obtain RG equations of soft SUSY breaking terms in
complicated models like models, 
which we shall discuss in the following sections.
Such procedure is as follows.
We replace $\alpha_g$ and $\alpha_y$ as \cite{Yamada:1994id}
\begin{eqnarray}
\alpha_g &\rightarrow & \tilde \alpha_g = \alpha_g ( 1 + M \theta^2  +
\bar M \bar \theta^2) + \cdots ,  \\
\alpha_y &\rightarrow & \tilde \alpha_y = \alpha_y ( 1 - A \theta^2  - 
\bar A \bar \theta^2) + \cdots ,
\end{eqnarray}
where $\theta $ is the Grassmann coordinate of the superspace 
and the ellipsis denotes terms with $\theta^2 \bar \theta^2$, 
which are irrelevant to the RG of the A-term, but 
relevant to soft scalar masses.
The couplings, $\tilde \alpha_g$ and $\tilde \alpha_y$, 
satisfy the same RG equations as those of $\alpha_g$ and $\alpha_y$.
That is, the RG equations of the gaugino mass $M$ and the A-term 
$A$ corresponding to eqs.(\ref{1-flavor-rg-g}) and
(\ref{1-flavor-rg-y}) are written as 
\begin{eqnarray}
\mu \frac{d M}{d \mu} &=& -b \alpha_g M, \\
\mu \frac{d A}{d \mu} &=& a \alpha_y A + c M \alpha_g,
\end{eqnarray}
where $a, b$ and $c$ are the same constants as eqs.(\ref{1-flavor-rg-g}) and
(\ref{1-flavor-rg-y}).
We can find that $\Delta A =(A+M)$ has a fixed point in these equation.
Actually, it satisfies the RG equation,
\begin{equation}
\mu \frac{d \Delta A}{d \mu} = - a \alpha_g M \Delta x 
+ a x^* \alpha_g \Delta A.
\end{equation}
At $x=x^*$, it reduces to 
\begin{equation}
\mu \frac{d \Delta A}{d \mu} = a x^* \alpha_g \Delta A
= (c-b) \alpha_g \Delta A.
\end{equation}
Note that the constant $a$ is always positive.
Thus, when the Yukawa coupling has the Pendelton-Ross 
IR fixed point, the corresponding A-term always has the fixed 
point, 
\begin{equation}
A = -M.
\label{1-flavor-M-A}
\end{equation}
The deviation from the fixed point $\Delta A$
decreases in the same way as $\Delta x$ satisfying 
eq.~(6).
Note that the gaugino mass $M$ and the A-term $A$ 
are complex parameters.
Thus, the above relation, $A= -M$, is realized 
including their CP phases, that is, 
the CP phase of the A-term is aligned with 
one of the gaugino mass at the IR fixed point.

\section{Infrared fixed points in three flavor model}

\subsection{Democratic fixed point}

In this section, we briefly review on the model 
with three flavors \cite{Kobayashi:2004ha}, in which the 
democratic form 
of the Yukawa matrix is realized dynamically by IR fixed point.

We consider the SUSY model including three flavors of 
matter superfields $F_i$ and $f_i$ $(i=1,2,3)$ and 
nine Higgs fields $H_{ij}$ with their superpotential,
\begin{equation}
W= \sum_{i,j}y_{ij} F_i f_j H_{ij}.
\end{equation}
Anomalous dimensions of these fields, 
$\gamma_{F_i}, \gamma_{f_j}$ and $\gamma_{H_{ij}}$, are 
written as 
\begin{eqnarray}
\gamma_{F_i} &=& a_F \sum_{k=1}^3 \alpha_{y_{ik}} - c_F \alpha_g, \\
\gamma_{f_j} &=& a_f \sum_{k=1}^3 \alpha_{y_{kj}} - c_f \alpha_g, \\
\gamma_{H_{ij}} &=& 3a_H \alpha_{y_{ij}} - c_H \alpha_g, 
\end{eqnarray}
where $\alpha_{y_{ij}} = |y_{ij}|^2/(8 \pi^2)$, 
$a_{F,f,H}$ are positive constants and $c_{F,f,H}$
are obtained as $c_{F,f,H}=2C_2(R_{F,f,H})$ by the corresponding 
quadratic Casimir $C_2(R_{F,f,H})$.
The RG equations of $\alpha_{y_{ij}}$ are given as 
\begin{equation}
\mu \frac{d \alpha_{y_{ij}}}{d \mu} = 
(\gamma_{F_i} + \gamma_{f_j} + \gamma_{H_{ij}}) \alpha_{y_{ij}}.
\end{equation}
As the previous section, we define 
$x_{ij} = \alpha_{y_{ij}}/\alpha_g$.
Then, it is found that the RG equations of $x_{ij}$ have 
the fixed point,
\begin{equation}
x_{ij} = x^* = \frac{c - b}{3a},
\end{equation}
where $a= a_F + a_f + a_H$.
To see whether this fixed point is IR attractive, 
we consider the linear perturbation around the fixed point, 
$x_{ij} = x^* + \Delta x_{ij}$.
The RG equations of $\Delta x_{ij}$ are obtained as 
\begin{equation}
\mu\frac{d}{d\mu}\Delta x_{ij}=3a_H \alpha_g x^{\ast}\Delta x_{ij}+
\sum_{k }a_F \alpha_g x^{\ast}\Delta x_{ik} +\sum_{k }a_f 
\alpha_g x^{\ast}\Delta x_{kj} .
\end{equation}
Alternatively, we can write more explicitly  
\begin{equation}
\mu \frac{d}{d\mu}\left(
\begin{array}{c} \Delta_1 \\ \Delta_2 \\ \Delta_3 \end{array}
\right)
=\alpha_g x^* \left(
\begin{array}{ccc} 
                    A & E & E \\
                    E & A & E \\
                    E & E & A
                 \end{array} \right)
\left(
\begin{array}{c} \Delta_1 \\ \Delta_2 \\ \Delta_3 \end{array}
\right) ,
\end{equation} 
where 
$\Delta_i=(\Delta x_{i1},\Delta x_{i2},\Delta x_{i3})^T$ and 
\begin{equation}
A=\left( \begin{array}{ccc} 
                    a' & a_F & a_F \\
                    a_F & a' & a_F \\
                    a_F & a_F & a'
                 \end{array}
\right) , \qquad 
E=\left( \begin{array}{ccc} 
                    a_f & 0 & 0 \\
                    0 & a_f & 0 \\
                    0 & 0 & a_f
                 \end{array}
\right) ,
\end{equation}
where $a' \equiv a_F +a_f +3a_H$.
This $(9 \times 9)$ matrix has the following eigenvalues,
\begin{equation}
3a_H,~3a_H,~3(a_F+a_H),~3a_H,~3a_H,~3(a_F+a_H),~3(a_f+a_H),~3(a_f+a_H),~3a .
\end{equation}
All of them are positive and 
it is found that the fixed point $x_{ij} = x^*$ is 
IR attractive.
Hence, this model dynamically realizes the 
democratic form of the Yukawa matrix, 
\begin{equation}
\alpha_{yij}^{\ast}\propto \left(
\begin{array}{ccc} 
                          1 & 1 & 1 \\
                          1 & 1 & 1 \\
                          1 & 1 & 1
                        \end{array}
\right) .
\end{equation}
To obtain the democratic form of the fermion mass matrix, 
we have to assume all of vacuum expectation values (VEVs) 
$\langle H_{ij} \rangle$ are the same.
Alternatively, we assume mass terms among $H_{ij}$ such that 
they lead to a single light mode $H=1/3\sum_{ij}H_{ij}$ and
the others have masses at a high energy scale.
Here we take the latter scenario, and such mass terms 
have been shown in Refs.\cite{Abel:1998wh,Kobayashi:2004ha}.
We denote their mass scale by $M_H$, and 
at $M_H$ the RG-flows approaching toward their fixed points are
terminated.

\subsection{Fixed points of A-terms}

Here, we consider fixed points of A-terms.
Using the spurion technique, we can write the RG equations of 
$A_{ij}$,
\begin{equation}
\mu \frac{d}{d \mu} A_{ij} = a_F \sum_k A_{ik}\alpha_{yik}+
a_f \sum_k A_{kj}\alpha_{ykj}+3a_HA_{ij}\alpha_{yij} +cM \alpha_g .      
\end{equation}
It is straightforward to show that there is the fixed point $A_{ij} = -M$.
Around the fixed points, we write the RG equations of 
$\Delta A_{ij} = A_{ij} +M$,
\begin{eqnarray}
\mu \frac{d}{d \mu}\Delta A_{ij} = 
&\alpha_g x^{*} \{ a_F \sum_k \Delta A_{ik} + a_f \sum_k \Delta A_{kj}
+3a_H\Delta A_{ij} \}   \nonumber \\
&-\alpha_g M \{ a_F \sum_k \Delta x_{ik} +a_f \sum_k \Delta x_{kj} +3a_H   \Delta x_{ij}         \}.
\label{RG-del-A}
\end{eqnarray}
When $\Delta x_{ij} = 0$, $\Delta A_{ij}$ satisfy the 
same RG equations as $\Delta x_{ij}$.
Thus, it is found that when the point $x_{ij} =x^*$ is 
IR attractive, the corresponding A-terms have the IR attractive 
fixed point,
\begin{equation}
A_{ij} = -M \left(
\begin{array}{ccc}
1 & 1 & 1\\
1 & 1 & 1\\
1 & 1 & 1
\end{array}
\right).
\end{equation}
This form of A-terms is quite important from the 
phenomenological viewpoint.
Obviously, the A-terms are universal, that is, 
favorable from the FCNC constraints.
Furthermore, the CP phase of the A-terms is 
aligned with the CP phase of the gaugino mass $M$, 
that is, there is no physical CP-violation phase 
in the A-terms.
Therefore, this form of A-terms can avoid 
both SUSY flavor and CP problems.
Constraints from FCNC processes as well as CP-violations 
are usually presented by using the mass 
insertion parameters, e.g. $(\delta^a_{LR})_{ij}$,
\begin{equation}
(\delta^a_{LR})_{ij} = \frac{y^a_{ij}A^a_{ij}}{m_{SUSY}^2}v^a,
\end{equation}
in the S-CKM basis, where $a=u,d,\ell$ and $m_{SUSY}$ 
denote the average of sfermion masses and $v^a$ denotes 
the VEV of the corresponding Higgs field.

\section{Realistic fermion masses}

\subsection{RG flow of Yukawa matrix}

In the previous section, we show the democratic form 
of the Yukawa matrix is realized at the IR fixed point, 
and at the same time the universal A-terms are 
obtained.
That is quite favorable from the viewpoint of 
SUSY flavor and CP problems.
However, the Yukawa matrix at the exact fixed point 
is not realistic, because it is a rank-one matrix, and 
only one family becomes massive and the other two 
remain massless.
A region deviated slightly from the fixed point 
would be interesting.
Thus, in this section we study the RG flows of 
the Yukawa matrix to investigate which pattern of 
the Yukawa matrix is obtained through a finite range of running
by generic initial conditions and examine the possibility 
for realistic Yukawa matrices.
In particular, we apply our analysis to the charged lepton sector.
In addition, in a such situation, the A-terms may also 
deviate from the universal form somehow.
We study whether such deviation leads to dangerous FCNCs.

The democratic matrix can be diagonalized as follows,
\begin{equation}
\frac{1}{3} \left(
\begin{array}{ccc}
1 & 1 & 1 \\
1 & 1 & 1 \\
1 & 1 & 1 
\end{array}
\right)
= U \left(
\begin{array}{ccc}
0 & 0 & 0 \\
0 & 0 & 0 \\
0 & 0 & 1 
\end{array}
\right)U^T,
\end{equation}
where 
\begin{equation}
U= \left(
\begin{array}{ccc}
\frac{1}{\sqrt{2}} & \frac{1}{\sqrt{6}} & \frac{1}{\sqrt{3}} \\
\frac{-1}{\sqrt{2}} & \frac{1}{\sqrt{6}} & \frac{1}{\sqrt{3}}  \\
0 & \frac{-2}{\sqrt{6}} & \frac{1}{\sqrt{3}} 
\end{array}
\right) .
\end{equation}

For example, 
if the lepton mixing matrix is obtained as 
$V_{MNS} \approx U^T$, that leads to almost realistic 
mixing angles, which are consistent with 
the neutrino oscillations \cite{Maltoni:2004ei}, i.e., 
\begin{equation}
(V_{MNS})_{12} \approx 0.5, \qquad 
(V_{MNS})_{23} \approx 0.7, \qquad (V_{MNS})_{13} < 0.1.
\end{equation}
That is the reason why the democratic form of 
the Yukawa matrix is interesting, in particular 
in the lepton sector.
That is, the Yukawa matrix of charged leptons may almost 
be democratic, and it determines the mixing angles dominantly
while the contribution from the neutrino Yukawa matrix 
to the mixing angle may be small \cite{Kobayashi:2004ha}.

Sometimes this ``hierarchical'' basis is more convenient to discuss 
than the democratic basis.
We denote the deviations of $x_{ij}$ from the fixed point in 
the hierarchical basis by $\Delta \tilde x_{ij}$.
Their RG equations are obtained as
\begin{equation}
\mu \frac{d}{d \mu} \Delta \tilde {x}_{ij}(\mu) = 
\alpha_g x^* \lambda_{ij} \Delta \tilde {x}_{ij} (\mu)  ,
\end{equation} 
with
\begin{equation}
\lambda_{ij} =\left(
\begin{array}{ccc}
 3a_H & 3a_H & 3(a_Q + a_H) \\
                          3a_H & 3a_H & 3(a_Q + a_H) \\
                          3(a_u +a_H) & 3(a_u+a_H) & 3a   
\end{array}
\right)  .
\end{equation}
Thus, they are solved as 
\begin{equation}
\Delta \tilde{x}_{ij} (\mu) =\Delta \tilde{x}_{ij} (\Lambda) 
\left( \frac{\alpha_g (\mu)}{\alpha_g (\Lambda)} 
\right)^{-\frac{x^*}{b} \lambda_{ij}} ,  
\end{equation}
and the Yukawa matrix in the hierarchical basis, 
$\tilde y_{ij}$  is 
obtained as 
\begin{equation}
\tilde y_{ij}  \sim 3\sqrt{x^*} g(\mu)   
\left(
\begin{array}{ccc}
\frac{\Delta \tilde{x}_{11} (\Lambda)}{6x^*} \varepsilon (\mu)^{3a_H} & 
\frac{\Delta \tilde{x}_{12} (\Lambda)}{6x^*} \varepsilon (\mu)^{3a_H} & 
\frac{\Delta \tilde{x}_{13} (\Lambda)}{6x^*} \varepsilon (\mu)^{3(a_Q+a_H)}  \\
\frac{\Delta \tilde{x}_{21} (\Lambda)}{6x^*} \varepsilon (\mu)^{3a_H} & 
\frac{\Delta \tilde{x}_{22} (\Lambda)}{6x^*} \varepsilon (\mu)^{3a_H} & 
\frac{\Delta \tilde{x}_{23} (\Lambda)}{6x^*} \varepsilon (\mu)^{3(a_Q+a_H)}  \\
\frac{\Delta \tilde{x}_{31} (\Lambda)}{6x^*} \varepsilon (\mu)^{3(a_u+a_H)} & 
\frac{\Delta \tilde{x}_{32} (\Lambda)}{6x^*} \varepsilon (\mu)^{3(a_u+a_H)} & 
1+ \frac{\Delta \tilde{x}_{33} (\Lambda)}{6x^*} \varepsilon (\mu)^{3a} 
\end{array} 
\right) ,
\end{equation}
where the suppression factor $\varepsilon(\mu)$ is 
obtained as 
\begin{equation}
\varepsilon(\mu) \equiv \left( 
\frac{\alpha_g(\mu)}{\alpha_g(\Lambda)} \right)^{-{x^*}/{b}}.
\end{equation}
Note that the powers of the suppression factor 
$\varepsilon(\mu)$ in the $(i,3)$ and $(3,i)$ entries 
are larger than those in the $(i,j)$ entries $(i,j=1,2)$.
Hence, when initial values $\Delta \tilde x_{ij}(\Lambda)$ 
are of the same order, e.g. of $O(1)$, 
the $(i,3)$ and $(3,i)$ entries are suppressed rapidly and 
we obtain the Yukawa matrix,
\begin{equation}
\tilde y_{ij} \sim 3\sqrt{x^*} g(\mu)   \left(
\begin{array}{ccc} 
\frac{\Delta \tilde{x}_{11} (\Lambda)}{6x^*} \varepsilon (\mu)^{3a_H} & 
\frac{\Delta \tilde{x}_{12} (\Lambda)}{6x^*} \varepsilon (\mu)^{3a_H} & 0  \\
\frac{\Delta \tilde{x}_{21} (\Lambda)}{6x^*} \varepsilon (\mu)^{3a_H} & 
\frac{\Delta \tilde{x}_{22} (\Lambda)}{6x^*} \varepsilon (\mu)^{3a_H} & 0  \\
 0 & 0 & 1 
\end{array}
\right) .
\label{leptonYukawa2}
\end{equation}
Thus, this RG-flow can explain why the third family is the heaviest.

To obtain numerically realistic result of the lepton sector, we need 
$\varepsilon (\mu)^{3a_H} \sim m_\mu / m_\tau$ when 
$\Delta \tilde x_{ij}/(6x^*) \sim 1$.
Furthermore, the $(2,3)$ mixing angle is almost determined 
by the (2,3) entry of $U^T$, because $\tilde y_{i3}$ are 
suppressed rapidly.
Thus, we obtain 
\begin{equation}
(V_{MNS})_{23} = \frac{2}{\sqrt{6}} ,
\end{equation}
up to a contribution from the neutrino sector.
This prediction is good when the contribution from 
the neutrino sector is not large \cite{Kobayashi:2004ha}.

On the other hand, the above form can not explain the 
mass hierarchy between the first and second families 
when all of initial values 
$\Delta \tilde x_{ij}(\Lambda)$ are of the same order, 
e.g. of $O(1)$.
If the initial values of the first and second families are 
chosen, e.g., 
\begin{equation}
\left(
\begin{array}{cc} 
\Delta \tilde{x}_{11}(\Lambda) & \Delta \tilde{x}_{12}(\Lambda) \\
\Delta \tilde{x}_{21}(\Lambda) & \Delta \tilde{x}_{22}(\Lambda) 
\end{array}
\right) = 
\Delta \tilde{x}(\Lambda)_{22}
\left(
\begin{array}{cc} 
 0 & \sqrt{\frac{m_e}{m_\mu}} \\
                \sqrt{\frac{m_e}{m_\mu}} & 1 
\end{array}
\right) ,
\label{lepton-initial}
\end{equation}
we obtain the correct mass ratio $m_e/m_\mu$.
In the democratic basis, these initial couplings are
given  as
\begin{equation}
\left(
\begin{array}{cc} 
\Delta x_{11}(\Lambda) & \Delta x_{12}(\Lambda) \\
\Delta x_{21}(\Lambda) & \Delta x_{22}(\Lambda) 
\end{array}
\right) = 
\Delta x(\Lambda)_{22}
\left(
\begin{array}{cc} 
 1 -  \sqrt{\frac{m_e}{m_\mu}} & 1 \\
1 & 1 + \sqrt{\frac{m_e}{m_\mu}} 
\end{array}
\right).
\label{lepton-initial2}
\end{equation}
In addition, it also contributes to the 
mixing angle $(V_{MNS})_{12}$ as 
\begin{equation}
(V_{MNS})_{12} \sim \frac{1}{\sqrt{2}}(1 - \sqrt{m_e/3m_\mu}),
\end{equation}
up to the contribution from the neutrino sector.
Indeed, the value $(U^T)_{12} = 1/\sqrt{2}$, which 
is derived from the exact democratic form,  is 
slightly different from the value consistent with 
the neutrino oscillations.
Thus, the above small modification is rather 
preferable.

\subsection{RG flow of A-terms}

Here, let us study the RG-flow of the corresponding 
A-terms.
In the hierarchical basis the RG equations (\ref{RG-del-A}) 
are written as 
\begin{equation}
\mu \frac{d \Delta \tilde A_{ij}}{d \mu} = 
\alpha_g x^{*} \lambda_{ij} \Delta \tilde A_{ij} - 
M \alpha_g \lambda_{ij} \Delta \tilde x_{ij}.
\end{equation}
Their solutions are obtained as 
\begin{equation}
\label{A-RG}
{\Delta} \tilde A_{ij}(\mu)=
\left[ \frac{\Delta \tilde A_{ij}(\Lambda)}{\Delta \tilde
    x_{ij}(\Lambda)} -\frac{\lambda_{ij} M(\Lambda)}{b} 
\left( 1 - \frac{\alpha_g(\mu)}{\alpha_g(\Lambda)} \right) 
\right] \Delta \tilde x_{ij} (\mu).
\end{equation}
Note that the damping factor is the same except the term 
including $\Delta \tilde x_{ij} (\mu)\alpha_g(\mu)/\alpha_g(\Lambda)$.

When we apply to the charged lepton sector, 
it is first noted that 
$\Delta \tilde A_{i3}$ and  $\Delta \tilde A_{3i}$
for $i,j=1,2$ are strongly suppressed  as $\tilde{y}_{ij}$
given in (\ref{leptonYukawa2}).
Therefore, the branching ratios for
the lepton flavor violating processes,
$\tau \rightarrow \mu \gamma$ and 
$\tau \rightarrow e \gamma$, turn out to be
much smaller than the experimental bounds.

It is found that another lepton flavor violating 
process, $\mu \rightarrow e \gamma$, is also
suppressed to some extent.
Since it is expected that 
$\varepsilon(\mu)^{3a_H} \sim (m_\mu/m_\tau)$,
we may evaluate 
\begin{eqnarray}
\Delta \tilde A_{ij} (\mu) &\sim&
(m_\mu/m_\tau)F_{ij}(\Lambda), \\
F_{ij}(\Lambda) &=& \Delta \tilde A_{ij} (\Lambda) - \frac{\lambda_{ij}
    M(\lambda)}{ b} \Delta \tilde x_{ij} (\Lambda),
\end{eqnarray}
for $i,j=1,2$.
Hence, the mass insertion parameters 
$\left( \delta^\ell_{LR} \right)_{ij}$ for $i,j = 1,2$
are estimated as 
\begin{equation}
\left( \delta^\ell_{LR} \right)_{ij} \sim (m_\mu/m_\tau)
(m_{\tau}/  m_{SUSY}) ,
\end{equation}
where $m_{SUSY}$ is the average mass of sleptons, and we have taken 
$F_{ij}(\Lambda) = m_{SUSY}$.
For example, in the case with $m_{SUSY} = 100$ GeV, 
we obtain 
$( \delta^\ell_{LR} )_{12} =10^{-4}$.
The second term in $F_{ij}(\Lambda)$ has more suppression 
factor like $O(10^{-1} -10^{-2})$ when we take 
$\Delta \tilde x_{12}(\Lambda) = \sqrt{m_e/m_\mu}$ like 
eq.~(\ref{lepton-initial}).
Furthermore, if $|b| = O(10)$, it would leads further
suppression factor by $O(10^{-1})$.
On the other hand, experimental bound on the $\mu \rightarrow e \gamma $ decay 
requires $( \delta^\ell_{LR} )_{12} \leq 10^{-6}$.
Furthermore, the experiment on the EDM of 
the electron requires $Im ( \delta^\ell_{LR} )_{11} \leq 10^{-7}$.
If there is no suppression factor for the first term 
$\Delta \tilde A_{ij} (\Lambda)$ in the $F_{ij}(\Lambda)$,
then these constraints are not satisfied. 
Conversely, if we have another mechanism to suppress 
$\Delta \tilde A_{ij} (\Lambda)$ \footnote{For example, 
a certain class of (string-motivated) supergravity 
models leads to $A_{ij}(\Lambda) = -M(\Lambda)$ as initial 
conditions. See e.g. Ref.~\cite{Kawamura:1997cw} and 
references therein.},
then the flavor non-universality generated through 
$\Delta \tilde{x}$ is sufficiently small.

\section{Realistic quark masses}

\subsection{RG flow of Yukawa couplings}

In the previous section, we studied a single sector 
of a Yukawa matrix, and apply to the charged lepton sector.
In this section, we study quark masses.
For quarks, both the up and down sectors of Yukawa coupling matrices 
should have democratic fixed points.
Otherwise, we can not realize small mixing angles.
Thus, we have to extend the previous analysis to 
the case including the up and down sectors of Yukawa matrices.

Here, we consider the concrete model, which has been proposed 
in Ref.~\cite{Kobayashi:2004ha}.
It is the $SU(5)' \times SU(5)''$ GUT model.
We assume three families of quarks as well as leptons correspond to 
${\bf 10}_i$ and $\bar {\bf 5}_i$ for $SU(5)'$ as the usual 
$SU(5)$ GUT, and they are singlets under $SU(5)''$.
We denote the gauge couplings $g'$ and $g''$ for 
$SU(5)'$ and $SU(5)''$, respectively.
We assume that $g'$ is strong, but $g''$ is weak.
We also assume that $SU(5)' \times SU(5)''$  is broken into 
the diagonal group $SU(5)$ at/above the GUT scale.
Then, we would obtain the usual $SU(5)$ GUT.
The gauge coupling $g$ of the diagonal group is obtained 
as $1/g^2 = 1/g'^2 + 1/g''^2$, and it is weak.
The Pendelton-Ross IR fixed points can be realized 
for $SU(5)'$ within short running, because 
the gauge coupling $g'$ is strong.
That is  one of the reasons why we extend the usual $SU(5)$ GUT 
to the $SU(5)' \times SU(5)''$ GUT.\footnote{
In extra dimensional models convergence toward IR 
fixed points may be sufficiently rapid \cite{Bando:2000it} 
in a single $SU(5)$ model.
Besides, the A-terms as well as the soft scalar masses are strongly
aligned into the universal for asymptotically free gauge theories
in the extra dimensions \cite{Kubo}.
}
We assume that the mass scale $M_H$ is around the GUT scale.

Another reason to consider the $SU(5)' \times SU(5)''$
GUT is related with the SUSY breaking parameters.
In the asymptotically non-free gauge theories considered
in the previous section, the running gaugino mass also
decreases towards lower energy scale.
Meanwhile the soft scalar masses are enhanced through the
strong gauge interaction and become much larger than the
gaugino mass.
In the extended GUT, however, there are two gaugino
masses $M'$ and $M''$ for the distinct gauge sectors of
$SU(5)'$ and $SU(5)''$ respectively, and the gaugino mass
$M$ obtained after the symmetry breaking is given by
\begin{equation}
\frac{M}{g^2} = \frac{M'}{g'^2} + \frac{M''}{g''^2}.
\end{equation}
Since we suppose $g'$ to be large, $M$ is almost the same
as $M''$, which is not reduced from it's initial value
due to smallness of the gauge coupling $g''$.
Therefore, the soft scalar masses, which may be enhanced
to be of the order of the initial $M''$,
do not dominate over the gaugino mass $M$, 
as long as the gaugino masses  $M'$ and $M''$ are given 
with their initial values of the same order.

We introduce nine pairs of Higgs fields 
$H^u({\bf 5})_{ij}$ and $H^d(\bar {\bf 5})_{ij}$, corresponding to 
the up and down sector Higgs fields.
We have the following superpotential;
\begin{equation}
W = y^u_{ij} {\bf 10}_i {\bf 10}_j H^u({\bf 5})_{ij} + 
y^d_{ij} {\bf 10}_i \bar {\bf 5}_j H^d(\bar {\bf 5})_{ij}.
\end{equation}
The anomalous dimensions of matter fields and Higgs fields are 
obtained as 
\begin{eqnarray}
\gamma_{{\bf 10}_i} &=& \sum_{k=1}^3 \{ a_Q^u(\alpha_{yik}^u +\alpha_{yki}^u)
+a_Q^d \alpha_{yik}^d \} -a_Q^u \alpha_{yii}^u -c_Q \alpha'_g , \\ 
\gamma_{\bar {\bf 5}_i} &=& a^d \sum_{k=1}^3 \alpha_{ykj}^d -c_d \alpha'_g ,\\
\gamma_{H^u_{ij}} &=& 3a_H^u  \alpha_{yij}^u -c_H^u \alpha'_g ,\\
\gamma_{H^d_{ij}} &=& 3a_H^d \alpha_{yij}^d -c_H^d \alpha'_g ,
\end{eqnarray}
with
$3a^u_H=6, \quad 3a^d_H=4, \quad a^u_Q=3, \quad a^d_Q=2, \quad
a^d=4, \quad 
c_Q=36/5, \quad c_d=c_H=24/5 $,
where $\alpha^{u,d}_{y_{ij}} = |y^{u,d}_{ij}|^2/(8 \pi^2)$ and 
$\alpha'_g = g'^2/(8 \pi^2)$.
Then, the RG equations of $\alpha^u_{y_{ij}},\alpha^d_{y_{ij}}$ 
are obtained as
\begin{eqnarray}
\mu \frac{d \alpha^u_{y_{ij}}}{d \mu} &=& 
\left (\gamma_{{\bf 10}_i} + \gamma_{{\bf 10}_j} +  \gamma_{H^u_{ij}} 
\right)  \alpha^u_{y_{ij}}, \\
\mu \frac{d \alpha^d_{y_{ij}}}{d \mu} &=& 
\left (\gamma_{{\bf 10}_i} + \gamma_{\bar {\bf 5}_j} +  
\gamma_{H^d_{ij}} 
\right)  \alpha^d_{y_{ij}} .
\end{eqnarray}

We consider the RG equations of 
$x_{ij}^{u,d} = {\alpha_{yij}^{u,d}}/{\alpha'_g}$, that is, 
\begin{eqnarray}
\mu \frac{d}{d \mu} x_{ij}^u &=& \{ (b'-2c_Q-c_H^u)+ 3a_H^u x_{ij}^u  
+\sum_{k=1}^3 a_Q^u(x_{ik}^u+x_{ki}^u+x_{kj}^u+x_{jk}^u ) \nonumber \\
& & +\sum_{k=1}^3 a_Q^d(x_{ik}^d+x_{jk}^d)  -
a_Q^u x_{ii}^u -a_Q^u x_{jj}^u \} \alpha'_g x_{ij}^u , \\
\mu \frac{d}{d \mu} x_{ij}^d &=& \{ (b'-c_Q-c_d-c_H^d)+ 3a_H^d x_{ij}^d  
\nonumber \\
                            &  &+\sum_{k=1}^3 a_Q^u(x_{ik}^u+x_{ki}^u
                            ) +\sum_{k=1}^3
                            a_Q^d(x_{ik}^d)+\sum_{k=1}^3 a^d
                            x_{kj}^d-a_Q^u x_{ii}^u \} 
\alpha'_g x_{ij}^d  ,
\end{eqnarray}
where $b'$ is the one-loop beta function coefficient of 
$g'$.
It is straightforward to show these RG equations have 
the following fixed points,
\begin{eqnarray}
x^u_{ij} &=& x^{u*}  =\frac{X(a_Q^d+a_H^d+a^d)-2a_Q^dY}{10a_Q^u(a_H^d +a^d)+3a_H^u(a_Q^d +a_H^d +a^d)}, \\
x^d_{ij} &=& x^{d*}=\frac{5a_Q^uX
  -(3a_H^u+10a_Q^u)Y}{30a_Q^da_Q^u-3(3a_H^u+10a_Q^u)(a_H^d
  +a_Q^d+a^d)} ,
\end{eqnarray}
where $X=2c_Q + c_H^u -b' $ and $Y=c_Q +c_d +c_H^d -b'$.
We consider perturbations around these fixed points, 
$\Delta_{ij}^u=x_{ij}^u-x^{u*}$ and $\Delta_{ij}^d=x_{ij}^d-x^{d*}$.
It is convenient to define 
\begin{equation}
\Delta_{ij}^S=\Delta_{ij}^u +\Delta_{ji}^u, \qquad 
\Delta_{ij}^A=\Delta_{ij}^u-\Delta_{ji}^u ,
\end{equation}
to solve their RG equations.
Actually, the RG equations of $\Delta_{ij}^{S,A,d}$ are written as 
\begin{eqnarray}
\mu \frac{d}{d \mu} \Delta_{ij}^A &=&3a_H^u x^{u*} \alpha'_g \Delta_{ij}^A ,\\
\mu \frac{d}{d \mu} \Delta_{ij}^S &=& \{ \sum_{k=1}^3 (
    2a_Q^u(\Delta^S_{ik}
    +\Delta^S_{kj})+2a_Q^d(\Delta^d_{ik}+\Delta^d_{jk})) 
\nonumber \\ & &+3a_H^u\Delta_{ij}^S-a_Q^u\Delta_{ii}^S
    -a_Q^u\Delta_{jj}^S \} \alpha'_g x^{u*}, \\
\mu \frac{d}{d \mu} \Delta_{ij}^d &=&\{ \sum_{k=1}^3 (
a_Q^u\Delta^S_{ik} +a_Q^d \Delta^d_{kj}+a^d \Delta^d_{kj} )
+3a_H^d\Delta_{ij}^d-\frac{a_Q^u}{2}\Delta_{ii}^S  \}\alpha'_g x^{d*}. 
\end{eqnarray}
Easily, we can solve the RG equations of $\Delta^A_{ij}$ as 
\begin{equation}
\Delta_{ij}^A (\mu) =\Delta_{ij}^A(\Lambda) \left(
  \frac{\alpha'_g(\mu)}{\alpha'_g(\Lambda)} \right)^{-{3a_H^ux^{u*}}/{b'}} .
\end{equation}
The other elements are mixed in the above equations.
Here, we use the hierarchical basis.
Through longsome but simple algebraic calculations, 
we find that the RG equations of 
$\tilde \Delta^S_{ij}$, $\tilde \Delta^d_{ij}$ and 
$\tilde \Delta^d_{3i}$ for $i,j=1,2$ are decoupled each other.
Then we can solve them,
\begin{eqnarray}
\tilde{\Delta}_{ij}^S (\mu) &=&\tilde{\Delta}_{ij}^S(\Lambda) 
\left( \frac{\alpha'_g(\mu)}{\alpha'_g(\Lambda)} 
\right)^{-{3a_H^ux^{u*}}/{b'}} ,\\
\tilde{\Delta}_{ij}^d (\mu) &=&\tilde{\Delta}_{ij}^d(\Lambda) 
\left( \frac{\alpha'_g(\mu)}{\alpha'_g(\Lambda)} 
\right)^{-{3a_H^dx^{d*}}/{b'}} ,\\
\tilde{\Delta}_{3i}^d (\mu) &=&\tilde{\Delta}_{3i}^d(\Lambda) 
\left( \frac{\alpha'_g(\mu)}{\alpha'_g(\Lambda)}
\right)^{-{3(a^d+a_H^d)
x^{d*}}/{b'}} ,
\end{eqnarray}
for $i,j= 1,2$.
However, the other elements $\tilde{\Delta}_{i3}^S$ and $\tilde{\Delta}_{i3}^d$
for $i=1,2,3$ are mixed as 
\begin{eqnarray} -b' \alpha'_g \frac{d}{d \alpha'_g} 
\left( \begin{array}{c} 
\tilde {\Delta}_{13}^S \\
  \tilde{\Delta}_{13}^d 
\end{array}  \right)
& =& 
\left( \begin{array}{cc} 
x^{u*}(3a_H^u+4a_Q^u) & 6a_Q^d x^{d*} \\ 
2x^{u*}a_Q^u &  3x^{d*}(a_Q^d+a_H^d)
\end{array} \right) 
\left(
\begin{array}{c}  \tilde{\Delta}_{13}^S \\ \tilde{\Delta}_{13}^d 
\end{array} \right)   \nonumber \\ 
& & -\frac{a_Q^ux^{u*}}{ \sqrt{2} } \left(
\begin{array}{c} 2 \\ 1 \end{array} \right) \tilde{\Delta}_{12}^S  , 
\end{eqnarray}   
\begin{eqnarray} -b' \alpha'_g \frac{d}{d \alpha'_g} \left(
\begin{array}{c} \tilde{\Delta}_{23}^S \\
  \tilde{\Delta}_{23}^d \end{array} \right) 
&=& \left( \begin{array}{cc}  x^{u*}(3a_H^u+4a_Q^u) & 6a_Q^d x^{d*} \\ 
2x^{u*}a_Q^u & 3x^{d*}(a_Q^d+a_H^d) \end{array} \right)  \left(
\begin{array}{c} \tilde{\Delta}_{23}^S \\ \tilde{\Delta}_{23}^d 
\end{array} \right)  \nonumber \\ 
& & -\frac{a_Q^ux^{u*}}{2 \sqrt{2} } \left(
\begin{array}{c} 2 \\ 1 \end{array} \right)
(\tilde{\Delta}_{11}^S-\tilde{\Delta}_{22}^S)  ,
\end{eqnarray}
\begin{eqnarray}   -b' \alpha'_g \frac{d}{d \alpha'_g} \left(
\begin{array}{c} \tilde{\Delta}_{33}^S \\
  \tilde{\Delta}_{33}^d \end{array} \right) 
&=& \left( \begin{array}{cc}  
x^{u*}(3a_H^u+10a_Q^u) & 12a_Q^d x^{d*} \\ 
\frac{5}{2}x^{u*}a_Q^u & 3x^{d*}(a_Q^d+A^d+a_H^d) \end{array} \right) 
\left( 
\begin{array}{c} \tilde{\Delta}_{33}^S \\
  \tilde{\Delta}_{33}^d \end{array} \right)  \nonumber \\ 
& & -\frac{a_Q^ux^{u*}}{2}
\left(
\begin{array}{c} 4 \\ 1 \end{array} \right)
(\tilde{\Delta}_{11}^S +\tilde{\Delta}_{22}^S)   .
\end{eqnarray}
The solutions $\tilde \Delta^S_{23}(\mu)$ and 
$\tilde \Delta^d_{23}(\mu)$ are obtained as 
\begin{eqnarray}\left(
\begin{array}{c} 
\tilde{\Delta}_{23}^S(\mu)/\tilde{\Delta}_{23}^S(\Lambda) \\ 
\tilde{\Delta}_{23}^d (\mu)/\tilde{\Delta}_{23}^d(\Lambda) 
\end{array} \right) &=& 
\left(
\begin{array}{cc} R_2 & Q_2 \\ R_2r_{+} & Q_2r_{-} \end{array} \right)
\left( \begin{array}{c} 
(\alpha'_g(\mu)/\alpha'_g(\Lambda))^{-n_+/b'} \\ 
(\alpha'_g(\mu)/\alpha'_g(\Lambda)))^{-n_-/b'} 
\end{array} \right) \nonumber \\
& & + \left( \begin{array}{c} k_1^2 \\ k_2^2 \end{array} \right)
(\alpha'_g(\mu)/\alpha'_g(\Lambda))^{-{3a_H^ux^{u*}}/{b'}},
\end{eqnarray}
where $n_\pm$ are obtained as 
\begin{equation}
n_\pm = \frac{D+A \pm \sqrt{(D-A)^2 +4BC} }{2},
\end{equation}
with 
\begin{equation}
\left( \begin{array}{cc}
A & B \\ C & D 
\end{array} \right) = \left( 
\begin{array}{cc}
x^{u*}(3a_H^u+4a_Q^u) & 6a_Q^d x^{d*} \\ 2x^{u*}a_Q^u &
  3x^{d*}(a_Q^d+a_H^d)
\end{array} \right) ,
\end{equation}
and similarly $r_\pm$ are obtained as 
\begin{equation}
r_\pm = \frac{D-A \pm \sqrt{(D-A)^2 +4BC} }{2B(-b')} .
\end{equation}
In addition, $R_i$ and $Q_i$ are integral constants 
determined by initial conditions.
Furthermore, the constants $k^2_1$ and $k^2_2$ are determined 
by solving the following equation;
\begin{equation}
\left(
\begin{array}{cc} l-A & -B \\ -C & l-D \end{array} 
\right)  \left(\begin{array}{c}  k^2_1 \\ k^2_2 \end{array}
\right) = -\frac{a_Q^ux^{u*}}{2 \sqrt{2}(-b') } \left(
\begin{array}{c} 2 \\ 1 \end{array} \right)
(\tilde{\Delta}_{11}^S-\tilde{\Delta}_{22}^S)  ,
\end{equation}
with $l=-{3 a_H^u x^{u*}}/{b'}$.
Substituting explicit values, we obtain 
\begin{equation}
n_{\pm} = 9x^{u*} + 5x^{d*} \pm \sqrt{(9x^{u*}-  5x^{d*})^2 + 
72 x^{u*} x^{d*}}.
\end{equation}
It is found that both $n_\pm$ are positive, and obviously 
$n_+ > n_-$.
The solutions $\tilde \Delta^S_{13}(\mu)$ and 
$\tilde \Delta^d_{13}(\mu)$ are the same as 
$\tilde \Delta^S_{23}(\mu)$ and 
$\tilde \Delta^d_{23}(\mu)$ except by replacing 
\begin{equation}
(\tilde{\Delta}_{11}^S-\tilde{\Delta}_{22}^S)  \rightarrow 
\tilde{\Delta}_{12}^S .
\end{equation}
Similarly we can solve $\tilde \Delta^S_{33}(\mu)$ and 
$\tilde \Delta^d_{33}(\mu)$, and they have rapidly damping 
behavior.
Since their explicit forms are irrelevant to our discussions, 
we omit them.
 
The solutions $\tilde \Delta^S_{i3}(\mu)$ and 
$\tilde \Delta^d_{i3}(\mu)$ for $i=1,2$ include 
three damping factors.
The most slowly damping term is important.
Since we always have $n_- < 6 (=3a^u_H)$, we find that 
$\tilde \Delta^S_{i3}(\mu)$ and 
$\tilde \Delta^d_{i3}(\mu)$ behave like $(\alpha'_g)^{-n_-/b'}$.
Thus, the up and down-sectors of Yukawa matrices behave totally as
\begin{equation}
y^u_{ij} \sim \left(
\begin{array}{ccc}
\varepsilon'(\mu)^{3a^u_Hx^{u*}} & \varepsilon'(\mu)^{3a^u_Hx^{u*}} & 
\varepsilon'(\mu)^{n_-} \\
\varepsilon'(\mu)^{3a^u_Hx^{u*}} & \varepsilon'(\mu)^{3a^u_Hx^{u*}} & 
\varepsilon'(\mu)^{n_-} \\
\varepsilon'(\mu)^{n_-} & \varepsilon'(\mu)^{n_-} & 1 
\end{array} \right), 
\end{equation}
\begin{equation}
y^d_{ij} \sim \left(
\begin{array}{ccc}
\varepsilon'(\mu)^{3a^d_Hx^{d*}} & \varepsilon'(\mu)^{3a^d_Hx^{d*}} & 
\varepsilon'(\mu)^{n_-} \\
\varepsilon'(\mu)^{3a^d_Hx^{d*}} & \varepsilon'(\mu)^{3a^d_Hx^{d*}} & 
\varepsilon'(\mu)^{n_-} \\
 0 & 0 & 1 
\end{array} \right), 
\end{equation}
where 
\begin{equation}
\varepsilon'(\mu) = \left( 
\frac{\alpha'_g(\mu)}{\alpha'_g(\Lambda)} \right)^{-1/b'} .
\end{equation}
We have omitted the coefficients.
Also we have omitted (3,1) and (3,2) entries in the down-sector 
Yukawa matrix, because they are damping more rapidly than 
the other entries in the down-sector Yukawa matrix.
That is the same behavior as in the case with a single Yukawa matrix, which 
discussed in the previous section.
However, $(1,3)$ and $(2,3)$ entries are damping not rapidly in 
the case including both the up and down-sector Yukawa matrices.
That would be important to derive the mixing angle 
$V_{cb}$ in the quark sector.

Recall that $3a^u_H =  6$ and $3a^d_H =  4$.
Explicit values of $n_-$ are $n_-= 3.5, 4.6$ and $5.3$ for 
$x^{u*} = 0.5 x^{d*}, x^{d*}$ and $2x^{d*}$, respectively.
Thus, when $x^{u*} \sim x^{d*}$, we can obtain the 
realistic relations,
\begin{equation}
\frac{m_c}{m_t} \sim \left( \frac{m_s}{m_b} \right)^{3/2}, 
\qquad V_{cb} \sim \frac{m_s}{m_b} .
\end{equation}
However, the model which we are discussing leads to
\begin{equation}
x^{u*} = \frac{92}{255}-\frac{5}{306}b', \qquad 
x^{d*} = \frac{44}{85}-\frac{7}{204}b' ,
\end{equation}
and it predicts rather large ratio of $m_c/m_t$ like 
$m_c/m_t \sim m_s/m_b$.
However, fixed point values $x^{u*}$ and $x^{d*}$ change 
significantly when we include other couplings.
In Appendix, we give examples, where fixed points 
are shifted by adding additional couplings.
In a such case, the above analysis is the same except 
by using new fixed point values $x^{u*}$ and $x^{d*}$ 
when additional couplings are close to their fixed points.
Thus, we could obtain models with several couplings leading 
to the realistic mass relations of the second and third 
families, e.g. the model leading to $x^{u*} \sim x^{d*}$.
Here we do not study explicitly a concrete model with 
values of $x^{u*}$ and $x^{d*}$ leading to realistic 
results.
 We use $x^{u*}$ and $x^{d*}$ as free parameters 
to present generic models with several couplings and 
effects due to such additional couplings.

We have shown that we can realize the quark 
mass ratios and the mixing angles between the second 
and third families when 
fixed point values $x^{u*}$ and $x^{d*}$ are 
in the proper region.
However, only the RG dynamics can not lead to 
the realistic mass hierarchy between 
the first and second families.
That is the same as the situation in the 
previous section.
Thus, we need fine-tuning of initial conditions.
For example, if the initial conditions are chosen, e.g., 
\begin{equation}
\left(
\begin{array}{cc} 
\tilde \Delta^u_{11} & \tilde \Delta^u_{12} \\
\tilde \Delta^u_{21} & \tilde \Delta^u_{22} 
\end{array}
\right) = 
\tilde \Delta^u (\Lambda)_{22}
\left(
\begin{array}{cc} 
 0 & \sqrt{\frac{m_u}{m_c}} \\
                \sqrt{\frac{m_u}{m_c}} & 1 
\end{array}
\right) ,
\end{equation}
\begin{equation}
\left(
\begin{array}{cc} 
\tilde \Delta^d_{11} & \tilde \Delta^d_{12} \\
\tilde \Delta^d_{21} & \tilde \Delta^d_{22} 
\end{array}
\right) = 
\tilde \Delta^d (\Lambda)_{22}
\left(
\begin{array}{cc} 
 0 & \sqrt{\frac{m_d}{m_s}} \\
                \sqrt{\frac{m_d}{m_s}} & 1 
\end{array}
\right) ,
\end{equation}
we obtain the correct mass ratios $m_u/m_c$ and 
$m_d/m_s$, and the mixing angle $V_{us} \sim m_d/m_s$.
These initial conditions are also consistent with 
those in the lepton sector, which was discussed 
in the previous section, because 
$m_e/m_\mu \sim m_d/m_s$.

\subsection{A-terms}

Here, we study the corresponding A-terms.
By using the spurion technique, we can obtain 
the RG equations of $A^{u,d}_{ij}$ and find 
they have the fixed points,
\begin{equation}
A_{ij}^{u,d} \rightarrow A^*=-M .
\end{equation}
We expand $A^{u,d}_{ij}$  around the fixed point as 
$A_{ij}^{u,d}=A^* +\Delta A_{ij}^{u,d}$, and 
the deviations $\Delta A_{ij}^{u,d}$ satisfy the 
following RG equations;
\begin{eqnarray}
\mu \frac{d}{d \mu} \Delta A_{ij}^S 
&=&\alpha'_g [ 3a_H^u x^{u*}\Delta A_{ij}^S -a_Q^u  x^{u*}\Delta A_{ii}^S -
a_Q^u  x^{u*}\Delta A_{jj}^S  \nonumber \\
& &+\sum_{k=1}^3 \{ 2a_Q^u  x^{u*}(\Delta A_{ik}^S + \Delta A_{kj}^S) 
+2a_Q^d  x^{d*}(\Delta A_{ik}^d +\Delta A_{jk}^d) \} ]  \nonumber \\
& &-\alpha'_g M [ 3a_H^u \Delta^S_{ij} -a_Q^u \Delta_{ii}^S-
a_Q^u \Delta_{jj}^S  \nonumber \\
& & +\sum_{k=1}^3 \{ 2a_Q^u (\Delta_{ik}^S + \Delta_{jk}^S) 
+2a_Q^d (\Delta_{ik}^d + \Delta_{jk}^d) \} ] ,\\
\mu \frac{d}{d \mu} \Delta A_{ij}^d 
&=&\alpha'_g [ 3a_H^d  x^{d*}\Delta A_{ij}^d -
\frac{a_Q^u }{2} x^{u*} \Delta A_{ii}^S    \nonumber \\
& &+\sum_{k=1}^3 \{ a_Q^u  x^{u*}\Delta A_{ik}^S  
+ (a_Q^d x^{d*}\Delta A_{ik}^d + a^d x^{d*}\Delta A_{kj}^d) \} ]  \nonumber \\
& &-\alpha'_g M \{ 3a_H^u \Delta^d_{ij} -\frac{a_Q^u}{2} \Delta_{ii}^S  
+\sum_{k=1}^3 ( a_Q^u \Delta_{ik}^S  +a_Q^d \Delta_{ik}^d 
+A^d \Delta_{jk}^d ) \} ,\\
\mu \frac{d}{d \mu} \Delta A_{ij}^A 
&=& 3a_H^u  x^{u*}\Delta A_{ij}^A \alpha'_g -\alpha'_g M 3a_H^u \Delta_{ij}^A ,
\end{eqnarray}
where $\Delta A_{ij}^S = \Delta A_{ij}^u + \Delta A_{ji}^u$ and 
$\Delta A_{ij}^A = \Delta A_{ij}^u - \Delta A_{ji}^u$.
Thus, we see that $\Delta A_{ij}$ behave like 
$\Delta A_{ij} \sim \Delta_{ij}$ similarly to the 
model in the previous section.

The experimental bounds for $\Delta M_K$, $\Delta M_B$
and the branching ratio of $b \rightarrow s \gamma$
restrict the mass insertion parameters as
$\left( \delta^{d}_{LR} \right)_{12} < 4.4 \times 10^{-3}$,
$\left( \delta^{d}_{LR} \right)_{13} < 3.3 \times 10^{-2}$
and   
$\left( \delta^{d}_{LR} \right)_{23} < 1.6 \times 10^{-2}$
respectively.
On the other hand, we can estimate the mass 
insertion parameters of the present model
$\left( \delta^{u,d}_{LR} \right)_{ij}$  
for $i,j= 1,2$ as 
\begin{equation}
\left( \delta^{u,d}_{LR} \right)_{ij} \sim (m_s/m_{SUSY}),
\end{equation}
where we have taken 
$\Delta A_{ij}(\Lambda) = m_{SUSY}$.
Then we obtain 
$\left( \delta^{u,d}_{LR} \right)_{12} = 10^{-4}$, 
which is consistent with the experiment on 
$\Delta M_K$.
We have taken $m_{SUSY} =500$GeV.
The mass insertion parameters 
$\left( \delta^{d}_{LR} \right)_{i3}$ 
for $i,j= 1,2$
are also found to be comparable with 
$\left( \delta^{d}_{LR} \right)_{12}$, 
since the parameter $n_-$ introduced in the previous
subsection is not so different from $3 a^d_H x^{d*}$.
Therefore, the bounds from  $\Delta M_B$
and $b \rightarrow s \gamma$ are also satisfied.

Moreover, the EDM of the neutron requires 
$Im \left (\left( \delta^{u,d}_{LR} \right)_{11} \right) = 10^{-6}$,
which seems to be rather severe in general.
However, this condition may be also explained,
when $|A_{ij}| \sim M'$ is smaller than 
$M'' \sim M$ by one or two orders at the GUT scale.
This is because the A-terms become almost flavor 
universal at the weak scale due to corrections through
the MSSM gauge interactions, which dominate over the 
initial values.
Otherwise, we need fine-tuning of 
initial conditions $\Delta \tilde A_{ij} (\Lambda)$ 
or some mechanism to lead to 
such required initial conditions.

\section{Comment on sfermion masses}

In the previous sections, we have concentrated 
to constraints on the A-terms.
Indeed, we have stronger constraints on the A-terms from 
FCNC and CP processes than soft sfermion masses.
Here we give a comment on sfermion masses. From the 
viewpoint of 
experiments on FCNC processes, 
degenerate sfermion masses are favorable.

The dynamics due to IR fixed points leads to 
the certain relation among soft scalar masses.
When the Yukawa coupling $y_{ij} F_i f_j H_{ij}$ has 
an IR fixed point, the sum of soft scalar masses 
squared 
also has the IR fixed point,
\begin{equation}
m_{\tilde F_i}^2 + m_{\tilde f_j}^2 + m_{H_{ij}}^2 = |M|^2.
\end{equation}
However, these relations are not enough to 
lead to degenerate sfermion masses among 
flavors when there are nine Higgs fields $H_{ij}$.
If there is any symmetry relating the soft scalar 
masses as well as the Yukawa couplings of the nine
Higgs fields, then the above relations
would lead to degenerate sfermion masses.
An example is the model with $A_4$ symmetry 
\cite{Babu:2002in}.
In this model, three families correspond to 
triplets under $A_4$, and four pairs of 
Higgs fields are introduced, and they 
correspond to $A_4$ singlet and triplet.
That is, three $H_{ii}$ are identified, i.e., 
$H_{11} = H_{22} =H_{33}$,  and 
$H_{ij}$ for $i \neq j$ is identified with $H_{ji}$, 
$H_{ij} = H_{ji}$, 
in words of our model with nine Higgs fields.
In such model, degeneracy of sfermion masses 
can be realized as IR fixed points \cite{Babu:2002ki}.  
However, in this model entries of the Yukawa matrix 
are also identified by the $A_4$ symmetry, 
that is, all of diagonal entries $y_{ii}$ are always the same, 
i.e., $y_{11}=y_{22}=y_{33}$,  and 
all of off-diagonal entries $y_{ij}$ ($i \neq j$) are always the same, 
i.e., $y_{ij} = y_{ji}$,  
even away IR fixed points.
We would obtain realistic results from such model when 
$y_{ii} = y_{ij}$ for $i \neq j$ and VEVs of 
four pairs of Higgs fields are fine-tuned in a proper 
way \cite{Fishbane:1993zc, Babu:2002in}.
Because that is a scenario different from the scenario 
in this paper, we do not study it here, but it may be 
another interesting scenario.

\section{Conclusion and discussion}

We have considered the model to realize the 
democratic form of Yukawa matrices.
We have studied the RG flows of Yukawa couplings and 
obtained their specific patterns.
The mass hierarchy and the mixing angles 
between the second and third families can be 
realized.
For a single sector of the Yukawa matrix, the 
mixing angle between the second and third families 
is determined almost by $U_{23}^T$.
However, for the up and down sectors of Yukawa matrices 
the mixing angle between the second and third families 
can be obtained as a value similar to $m_s/m_b$, i.e., 
$m_s/m_b \sim V_{cb} $.
These aspects are quite interesting.
Also, we can obtain the mass hierarchy, 
$m_s/m_b \sim m_{\mu}/m_\tau$ and 
$m_c/m_t \sim (m_s/m_b)^{3/2}$.

The corresponding A-term couplings also have 
the universal IR fixed point.
That is also important in order to 
avoid SUSY FCNC and CP problems.
In particular, there are strong constraints from 
the Kaon system, 
the $\mu \rightarrow e \gamma$ and EDMs.
FCNC constraints can be relaxed in the parameter region 
realizing realistic values of quark mass hierarchies, 
except for the $\mu \rightarrow e \gamma$ process.
Constraints due to CP violations can be ameliorated, 
but that may not be sufficient.
We would need fine-tuning of the initial conditions 
on the A-terms.
Alternatively, we may need some mechanism to 
realize such fine-tuning.

Our model does not intend to explain the mass 
hierarchy and  mixing angle between the first 
and second families.
In order to do this, we need to choose proper initial
conditions for them, or to find some mechanism to 
realize such initial conditions.
For example, we may consider to achieve dynamical
alignment of Yukawa couplings so that the large mass hierarchy 
like $m_u/m_t$ 
can be generated \cite{Kobayashi:2004ha}.
This can be realized by considering a strongly 
coupled GUT.
Moreover, it was shown in Ref.~\cite{Kobayashi:2004ha}
that the mass matrix close to the Fritzsch-type
one can be obtained, if one of the Yukawa couplings
is given with a relatively small initial value. 
In such a situation, the gauge dynamics aligns
the A-terms strongly enough, and the problematic
processes of FCNC and CP violations may be
automatically suppressed well below their 
experimental bounds.
That is beyond the scope of this paper and 
we would study it elsewhere.

Within the framework of our model with nine pairs 
of Higgs fields, we can not control squark/slepton 
masses only by fixed point dynamics 
such that they are degenerate to avoid FCNC constraints.

\section*{Acknowledgements}

T.~K.\ is supported in part by the Grants-in-Aid for 
Scientific Research
(No.~17540251) 
and the Grant-in-Aid for the 21st Century COE
``The Center for Diversity and Universality in Physics''
from the Ministry of Education, Science, Sports and 
Culture, Japan.
H.~T.\ 
is also supported in part by the Grants-in-Aid for 
Scientific Research No.~40192653
from the Ministry of Education, Science, Sports and 
Culture, Japan.

\section*{Appendix}

Here we give two examples of models with additional couplings, 
which shift values of fixed points.

First, we consider the model, which has the coupling among 
$H^{u,d}_{ij}$ and the adjoint field ${\bf 24}$.
The relevant superpotential is written as 
\begin{equation}
W=y_{ij}^u {\bf 10}^i {\bf 10}^j H^u({\bf 5})_{ij} +y_{ij}^d {\bf 10}^i
\bar {\bf 5}^j H^d(\bar {\bf 5})_{ij}
+\kappa_{ij,kl}H^u({\bf 5})^{ij} H^d(\bar {\bf 5})^{kl}{\bf 24} .
\end{equation}
There may be other couplings as well as other fields.
However, we assume that only the above couplings are 
around their fixed points, 
and we neglect other couplings.
In this model, anomalous dimensions are obtained as 
\begin{eqnarray}
\gamma_{{\bf 10}_i} &=&\sum_{k=1}^3 \{ a_Q^u(\alpha_{y_{ik}}^u 
+\alpha_{y_{ki}}^u) 
+a_Q^d \alpha_{y_{ik}}^d \} -a_Q^u \alpha_{y_{ii}}^u -c_Q \alpha'_g  ,\\ 
\gamma_{\bar{\bf 5}_i} &=&a^d \sum_{k=1}^3 \alpha_{y_{kj}}^d -c_d \alpha'_g ,\\
\gamma_{H^u_{ij}} &=&3a_H^u  \alpha_{y_{ij}}^u 
+d_H \sum_{k,l} \alpha_{ij,kl}-c_H^u \alpha'_g ,\\
\gamma_{H^d_{ij}} &=&3a_H^d \alpha_{y_{ij}}^d 
+d_H \sum_{k,l} \alpha_{kl,ij} -c_H^d \alpha'_g  ,\\
\gamma_{\bf 24} &=& a_{\bf 24} \sum_{ij,kl} \alpha_{ij,kl} -c_{\bf 24} 
\alpha'_g ,
\end{eqnarray}
where $a_{\bf 24}=1$, $c_{\bf 24} = 10$, $d_H=5$ and 
$\alpha_{ij,kl} = |\kappa_{ij,kl}|^2/(8\pi^2)$.
The RG equations of Yukawa couplings are written as 
\begin{eqnarray}
\mu \frac{d}{d \mu} \alpha_{y_{ij}}^u &=& ( \gamma_{{\bf 10}_i} +
\gamma_{{\bf 10}_j} +\gamma_{H^u_{ ij}})\alpha_{y_{ij}}^u   ,\\
\mu \frac{d}{d \mu} \alpha_{y_{ij}}^d &=& ( \gamma_{{\bf 10}_i} 
+\gamma_{\bar{\bf 5}_j} +\gamma_{H^d _{ij}})\alpha_{y_{ij}}^d   ,\\
\mu \frac{d}{d \mu} \alpha_{ij,kl} &=& (\gamma_{H^u_{ ij}} 
+\gamma_{H^d_{  kl}} +\gamma_{\bf 24}) \alpha_{ij,kl} .
\end{eqnarray}
This model has the IR fixed points,
\begin{eqnarray}
   x^{u*} &=& \frac{3959 -175 b'}{13635} = 0.29 - 0.013 b' ,\\
   x^{d*} &=& \frac{3361 -245 b'}{9090}  = 0.37 - 0.027 b',
\end{eqnarray}

Next we consider the model, which includes extra fields 
${\bf 5}'$ and it couples with $H^d_{ij}$ and 
the adjoint field.
Other extra fields are added such that this model is anomaly-free.
The relevant superpotential is written as 
\begin{equation}
W=y_{ij}^u {\bf 10}^i {\bf 10}^j H^u({\bf 5})_{ij} +y_{ij}^d {\bf 10}^i
\bar {\bf 5}^j H^d(\bar {\bf 5})_{ij}
+\kappa'_{kl}{\bf 5}' H^d(\bar {\bf 5})^{kl}{\bf 24} .
\end{equation}
The anomalous dimensions are obtained as 
\begin{eqnarray}
\gamma_{{\bf 10}_i}&=&\sum_{k=1}^3 \{ a_Q^u(\alpha_{y_{ik}}^u 
+\alpha_{y_{ki}}^u) 
+a_Q^d \alpha_{y_{ik}}^d \} -a_Q^u \alpha_{y_{ii}}^u -c_Q \alpha'_g , \\ 
\gamma_{\bar{\bf 5}_i} &=&a^d \sum_{k=1}^3 \alpha_{y_{kj}}^d -c_d \alpha'_g ,\\
\gamma_{H^u_{ij}} &=&3a_H^u  \alpha_{y_{ij}}^u  -c_H^u \alpha'_g ,\\
\gamma_{H^d_{ij}} &=&3a_H^d \alpha_{yij}^d +d_H \alpha'_{ij}  -c_H^d \alpha'_g  \\
\gamma_{\bf 24} &=& d_{\bf 24} \sum_{ij} \alpha'_{ij} -c_{\bf 24} \alpha'_g ,\\
\gamma_{{\bf 5}'} &=& d_H \sum_{ij} \alpha'_{ij} -c_H \alpha'_g, 
\end{eqnarray}
where $\alpha'_{ij} =|\kappa_{ij}|^2/(8\pi^2)$.
We can obtain the RG equations of Yukawa couplings.
Then we find that they have the IR fixed points,
\begin{eqnarray}
   x^u &=& \frac{12828 -575 b'}{27270} = 0.47 - 0.021 b' ,\\
   x^d &=& \frac{3432 -365 b'}{18180}  = 0.19 - 0.020 b' .
\end{eqnarray}
Thus, as seen in these two models, values of fixed points 
$x^{u*}$ and $x^{d*}$ are shifted significantly.



\begin{thebibliography}{99}


\bibitem{Fritzsch:1979zq}
  H.~Fritzsch,
  Nucl.\ Phys.\ B {\bf 155}, 189 (1979);
%
  Y.~Koide,
  Phys.\ Rev.\ D {\bf 28}, 252 (1983);
%
  Phys.\ Rev.\ D {\bf 39}, 1391 (1989).

\bibitem{Fritzsch:1995dj}
  H.~Fritzsch and Z.~Z.~Xing,
  Phys.\ Lett.\ B {\bf 372}, 265 (1996)
  [arXiv:hep-ph/9509389];
%
%
  Phys.\ Lett.\ B {\bf 440}, 313 (1998)
  [arXiv:hep-ph/9808272];
%
  Phys.\ Rev.\ D {\bf 61}, 073016 (2000)
  [arXiv:hep-ph/9909304];
%
  Phys.\ Lett.\ B {\bf 598}, 237 (2004)
  [arXiv:hep-ph/0406206].

\bibitem{Tanimoto:2000fz}
  M.~Tanimoto,
  Phys.\ Lett.\ B {\bf 483}, 417 (2000)
  [arXiv:hep-ph/0001306];
%
  M.~Fukugita, M.~Tanimoto and T.~Yanagida,
  Phys.\ Rev.\ D {\bf 57}, 4429 (1998)
  [arXiv:hep-ph/9709388].

\bibitem{Tanimoto:1999pj}
  M.~Tanimoto, T.~Watari and T.~Yanagida,
  Phys.\ Lett.\ B {\bf 461}, 345 (1999)
  [arXiv:hep-ph/9904338];
%
  E.~K.~Akhmedov, G.~C.~Branco, F.~R.~Joaquim and J.~I.~Silva-Marcos,
  Phys.\ Lett.\ B {\bf 498}, 237 (2001)
  [arXiv:hep-ph/0008010];
%
  G.~C.~Branco and J.~I.~Silva-Marcos,
  Phys.\ Lett.\ B {\bf 526}, 104 (2002)
  [arXiv:hep-ph/0106125];
%
  T.~Watari and T.~Yanagida,
  Phys.\ Lett.\ B {\bf 544}, 167 (2002)
  [arXiv:hep-ph/0205090];
%
  Q.~Shafi and Z.~Tavartkiladze,
  Phys.\ Lett.\ B {\bf 594}, 177 (2004)
  [arXiv:hep-ph/0401235].





\bibitem{Kobayashi:2004ha}
  T.~Kobayashi, H.~Shirano and H.~Terao,
Prog.~Thoer.~Phys. {\bf 113} (2005) 1077  
  [arXiv:hep-ph/0412299].


\bibitem{Abel:1998wh}
  S.~A.~Abel and S.~F.~King,
  Phys.\ Lett.\ B {\bf 435}, 73 (1998)
  [arXiv:hep-ph/9804446].


\bibitem{Pendleton:1980as}
  B.~Pendleton and G.~G.~Ross,
  Phys.\ Lett.\ B {\bf 98}, 291 (1981).

\bibitem{Gabbiani:1996hi}
F.~Gabbiani, E.~Gabrielli, A.~Masiero and L.~Silvestrini,
B {\bf 477}, 321 (1996)
[arXiv:hep-ph/9604387].

\bibitem{Chankowski:2005jh}
P.~H.~Chankowski, O.~Lebedev and S.~Pokorski,
arXiv:hep-ph/0502076.

\bibitem{Hamaguchi}
K.~Hamaguchi, M.~Kakizaki and M.~Yamaguchi,
Phys.~Rev.~{\bf D68}, 056007 (2003)
[arXiv:hep-ph/0212172].

\bibitem{Lanzagorta:1995ai}
  M.~Lanzagorta and G.~G.~Ross,
  Phys.\ Lett.\ B {\bf 364}, 163 (1995)
  [arXiv:hep-ph/9507366];
%
  P.~M.~Ferreira, I.~Jack and D.~R.~T.~Jones,
  Phys.\ Lett.\ B {\bf 357}, 359 (1995)
  [arXiv:hep-ph/9506467].





\bibitem{Kobayashi:2000wk}
  T.~Kobayashi and K.~Yoshioka,
  Phys.\ Lett.\ B {\bf 486}, 223 (2000)
  [arXiv:hep-ph/0004175];
%
  T.~Kobayashi and H.~Terao,
  Phys.\ Lett.\ B {\bf 489}, 233 (2000)
  [arXiv:hep-ph/0005265].


\bibitem{Kobayashi:2000fi}
  T.~Kobayashi and K.~Yoshioka,
  Phys.\ Rev.\ D {\bf 62}, 115003 (2000)
  [arXiv:hep-ph/0005009].


\bibitem{Yamada:1994id}
  Y.~Yamada,
  Phys.\ Rev.\ D {\bf 50}, 3537 (1994)
  [arXiv:hep-ph/9401241];
%
  J.~Hisano and M.~A.~Shifman,
  Phys.\ Rev.\ D {\bf 56}, 5475 (1997)
  [arXiv:hep-ph/9705417];
%
  I.~Jack and D.~R.~T.~Jones,
  Phys.\ Lett.\ B {\bf 415}, 383 (1997)
  [arXiv:hep-ph/9709364];
%
  I.~Jack, D.~R.~T.~Jones and A.~Pickering,
  Phys.\ Lett.\ B {\bf 426}, 73 (1998)
  [arXiv:hep-ph/9712542];
%
  Phys.\ Lett.\ B {\bf 432}, 114 (1998)
  [arXiv:hep-ph/9803405];
%
  L.~V.~Avdeev, D.~I.~Kazakov and I.~N.~Kondrashuk,
  Nucl.\ Phys.\ B {\bf 510}, 289 (1998)
  [arXiv:hep-ph/9709397];
%
  T.~Kobayashi, J.~Kubo and G.~Zoupanos,
  Phys.\ Lett.\ B {\bf 427}, 291 (1998)
  [arXiv:hep-ph/9802267];
%
  N.~Arkani-Hamed, G.~F.~Giudice, M.~A.~Luty and R.~Rattazzi,
  Phys.\ Rev.\ D {\bf 58}, 115005 (1998)
  [arXiv:hep-ph/9803290];
%
  D.~I.~Kazakov and V.~N.~Velizhanin,
  Phys.\ Lett.\ B {\bf 485}, 393 (2000)
  [arXiv:hep-ph/0005185];
%
  H.~Terao,
  arXiv:hep-ph/0112021.


\bibitem{Maltoni:2004ei}
  M.~Maltoni, T.~Schwetz, M.~A.~Tortola and J.~W.~F.~Valle,
  New J.\ Phys.\  {\bf 6}, 122 (2004)
  [arXiv:hep-ph/0405172].






\bibitem{Kawamura:1997cw}
  Y.~Kawamura, T.~Kobayashi and J.~Kubo,
  Phys.\ Lett.\ B {\bf 405}, 64 (1997)
  [arXiv:hep-ph/9703320].




\bibitem{Bando:2000it}
  M.~Bando, T.~Kobayashi, T.~Noguchi and K.~Yoshioka,
  Phys.\ Lett.\ B {\bf 480}, 187 (2000)
  [arXiv:hep-ph/0002102];
  Phys.\ Rev.\ D {\bf 63}, 113017 (2001)
  [arXiv:hep-ph/0008120].

\bibitem{Kubo}
J.~Kubo and H.~Terao, Phys.\ Rev.\ D {\bf 66}, 116003 (2002);
Y.~Kajiyama, J.~Kubo and H.~Terao,
Phys.\ Rev.\ D {\bf 69}, 116006 (2004)
[arXiv:hep-ph/0311316];
K-Y.~Choi, Y.~Kajiyama, H.M.~Lee and J.~Kubo,
Phys.\ Rev.\ D {\bf 70}, 055004 (2004)
[arXiv:hep-ph/0402026].

\bibitem{Babu:2002in}
  K.~S.~Babu, T.~Enkhbat and I.~Gogoladze,
  Phys.\ Lett.\ B {\bf 555}, 238 (2003)
  [arXiv:hep-ph/0204246].


\bibitem{Babu:2002ki}
  K.~S.~Babu, T.~Kobayashi and J.~Kubo,
  Phys.\ Rev.\ D {\bf 67}, 075018 (2003)
  [arXiv:hep-ph/0212350].

\bibitem{Fishbane:1993zc}
  P.~M.~Fishbane and P.~Kaus,
  Phys.\ Rev.\ D {\bf 49}, 4780 (1994);
  Z.\ Phys.\ C {\bf 75} (1997) 1.





\end{thebibliography}
\end{document}